\documentclass[final,namedreferences,hyperref,optionalrh]{spr-sola}

\usepackage{graphicx}
\usepackage{color}

\usepackage{amsmath}
\usepackage{amssymb}
\usepackage{bm}

\usepackage{float}
\usepackage{placeins}

\usepackage{booktabs}
\usepackage{multirow}
\usepackage{array}

\usepackage{tabularx}

\chardef\us=`\_

\hypersetup{
  colorlinks=true,
  linkcolor=black,
  citecolor=black,
  urlcolor=black
}

\usepackage[a4paper, top=2.5cm, bottom=2.5cm, left=2.5cm, right=2.5cm]{geometry}

\begin{document}

\begin{frontmatter}
\title{X-CME: From In Situ Flux-Rope Reconstruction to CME Propagation Forecasting}

\author[addressref={aff1,aff2},corref,email={marti.masso.moreno@estudiantat.upc.edu}]{\inits{M.}\fnm{Martí}~\snm{Massó~Moreno}}
\author[addressref={aff2,aff3},email={cperezal@gmu.edu}]{\inits{C.A.}\fnm{Carlos~Arturo}~\snm{Pérez-Alanis}}
\author[addressref={aff2,aff4},email={manoharanp@cua.edu}]{\inits{P.~K.}\fnm{P.~K.}~\snm{Manoharan}\orcid{0000-0003-4274-211X}}

\address[id=aff1]{CFIS, Universitat Politècnica de Catalunya (UPC), Barcelona, Spain}
\address[id=aff2]{Heliophysics Science Division, NASA Goddard Space Flight Center, Greenbelt, MD 20771, USA}
\address[id=aff3]{George Mason University, Fairfax, VA 22030, USA}
\address[id=aff4]{The Catholic University of America, Washington, DC 20664, USA}

\runningauthor{Massó~Moreno et al.}
\runningtitle{X-CME: From In Situ Flux-Rope Reconstruction to CME Propagation Forecasting}

\begin{abstract}
Accurate forecasts of Coronal Mass Ejection (CME) arrival times and impact geometry remain a major challenge for space-weather operations. Coronagraph-based techniques typically achieve mean absolute errors of order ten hours, while in situ measurements at L1 provide excellent magnetic-field information but only tens of minutes of warning. In this work we introduce X-CME, a framework that links in situ flux-rope reconstructions at intermediate heliocentric distances with a physics-based CME–propagation model. The internal magnetic structure is obtained with an elliptical–cylindrical, radial–poloidal flux-rope model and embedded into a tapered-torus CME geometry. The subsequent propagation is computed by solving a drag-based equation of motion in a Parker solar-wind background, including gravitational deceleration, self-similar expansion of the cross-section, and an explicit calculation of the CME wetted area and swept area in the ecliptic plane. We apply X-CME to two events observed by \emph{Parker Solar Probe} and \emph{Solar Orbiter}, respectively, and propagate the reconstructed structures to Earth and Mars. For both cases, the model reproduces the observed in situ signatures at L1 and predicts the CME arrival time at Earth with errors of a few hours (typically $\sim$2--4~h), while correctly distinguishing between central encounters and glancing blows. These results demonstrate that combining intermediate-distance magnetic reconstructions with a geometrically consistent propagation scheme can substantially improve CME arrival-time forecasts and impact assessment in the inner heliosphere.
\end{abstract}


\end{frontmatter}

\section{Introduction}
\label{S-Introduction}

Forecasting the arrival of Coronal Mass Ejections (CMEs) at Earth remains a critical challenge in space-weather prediction. Traditional methods rely heavily on coronagraph-based observations, which use sequences of white-light images to reconstruct the three-dimensional geometry and kinematics of CMEs as they erupt from the solar corona. Techniques based on geometric triangulation or forward modeling have provided valuable insights into CME trajectories and speeds; however, arrival-time predictions at 1~AU often suffer from mean absolute errors of order ten hours \citep{2024SpWea..2203951K}. These inaccuracies arise from projection effects, limited temporal cadence of coronagraph imagers, and simplifying assumptions regarding CME expansion, morphology, and interaction with the solar wind. The problem becomes even more acute when CMEs undergo significant non-radial deflections, rotations, or interact with background solar-wind structures, all of which are difficult to capture using coronagraph imagery alone.

In contrast, in situ detection at L1 provides direct, high-resolution measurements of plasma density, bulk speed, and magnetic-field vectors as the Interplanetary CME (ICME) reaches near-Earth space. Spacecraft such as ACE, WIND, and DSCOVR deliver precise characterization of the internal magnetic structure, enabling detailed reconstructions of the Magnetic Flux Rope (MFR). Nevertheless, the warning time provided by L1 is inherently limited: once the leading edge of the ICME crosses the spacecraft, only 15–30 minutes typically remain before impact at Earth \citep{2018ARA&A..56..135Z}, offering insufficient time to implement meaningful mitigation strategies for satellites, power grids, or crewed space missions.

The present work proposes an intermediate approach that bridges the temporal and spatial gap between early coronagraph detections and late-stage L1 measurements. By exploiting heliocentric spacecraft at intermediate distances—such as \emph{Parker Solar Probe} (PSP) at $\sim\!0.25$~AU and \emph{Solar Orbiter} (SolO) at $\sim\!0.6$~AU—it becomes possible to capture the evolving CME structure well before it reaches Earth. Specifically, we reconstruct the internal magnetic configuration of the CME’s flux rope using in situ data from these spacecraft and subsequently propagate the reconstructed structure through the remaining heliocentric distance to predict arrival times at Earth. This strategy combines the early-detection capabilities of remote sensing with the magnetic-detail advantages of in situ sampling, providing a pathway toward more accurate and operationally useful space-weather forecasts.

\section{CME Propagation Model}

\subsection{Model Considerations}

The reconstruction model of the CME begins with a local flux–rope reconstruction using the elliptical–cylindrical (EC) model presented in \citep{Nieves18}, further generalized in a companion manuscript \citep{MasoMoreno2025prep}. The locally reconstructed cross–section and magnetic configuration are then embedded into a global Tapered Torus geometry, with poloidal angle $\phi$ spanning from $-\pi/2$ to $\pi/2$ (croissant–shaped torus). The following assumptions are adopted throughout this work:

\begin{itemize}
    \item The in situ measured velocity of the flux rope is assumed to be predominantly radial, implying nearly radial propagation of the CME away from the Sun (Mishra \textit{et~al.}, \citeyear{Mishra2021}; Kay \textit{et~al.}, \citeyear{Kay2015}).
    
    \item The total CME mass is taken to remain constant during interplanetary propagation, consistent with coronagraphic and in situ studies showing that CME mass estimates stabilize beyond $\sim\!10$--15~$R_{\odot}$ with negligible subsequent accumulation \citep{Vourlidas2010, Bein2013}.
    
    \item The CME plasma is treated as globally electrically neutral, in accordance with the quasi–neutrality of space plasmas, where any charge imbalance is rapidly neutralized on scales far smaller than the CME size \citep{MeyerVernet2007, Chen2016}.

\begin{figure}[!ht]
    \centering
    \includegraphics[width=0.9\linewidth]{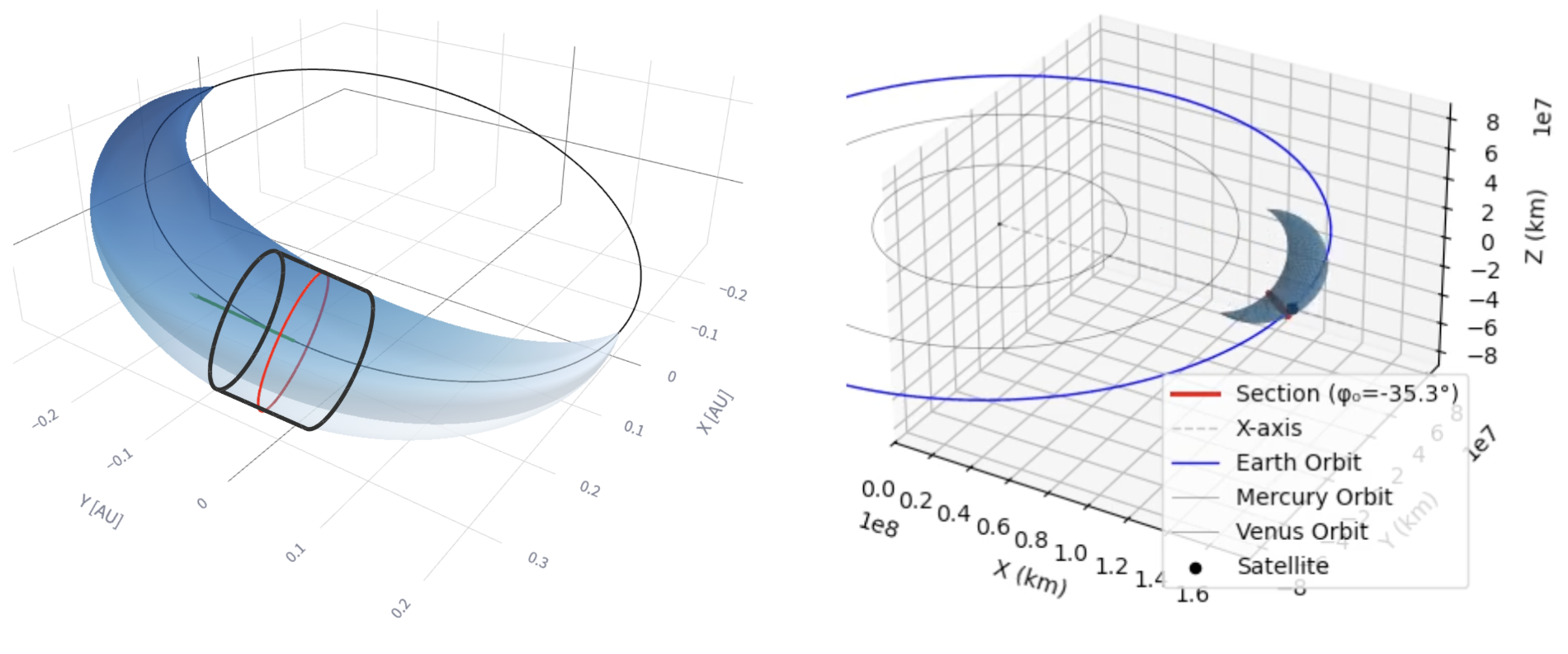}
    \caption{a) Local cylindrical approximation based on the EC flux–rope reconstruction. 
    b) Orientation of the CME and alignment of the cross–section with the Sun–spacecraft direction.}
    \label{fig:placeholder}
\end{figure}

    \item Local approximations: following the EC flux–rope formalism, we assume local axial invariance of physical quantities ($\partial_y \square = 0$, where $\square$ denotes magnetic field, current density, etc.). A locally cylindrical geometry is also assumed, allowing the EC reconstruction to be consistently embedded into the global Tapered Torus.
    
    \item The CME cross–section is assumed constant along the structure and identical in shape to the flux–rope cross–section inferred from the in situ reconstruction.
    
    \item For the flux–rope fitting, the ends of the Tapered Torus are excluded to avoid regions where the local cylindrical approximation breaks down; margins of $45^{\circ}$ are left at each CME leg.
    
    \item The CME expands self–similarly with heliocentric distance, following a power–law profile calibrated from the observed cross–sectional sizes at the spacecraft location (see Section~\ref{Mass_and_ProjectedArea}).
\end{itemize}

\subsection{Dynamics of the CME Propagation}

As shown in \citep{Vrsnak10}, the primary contribution to the dynamics of a CME during its propagation is the drag due to the solar wind. However, this holds true beyond 20 solar radii. Before that, the gravitational interaction with the Sun is also relevant. Magnetic interaction is very significant at distances closer to the Sun, but for this model, we will not consider it. Therefore, our system takes the following form:

\begin{equation}
m_{\text{CME}} \cdot a_{\text{CME}} = F_{\text{grav}} + F_{\text{drag}}
\end{equation}

which can be rewritten as:

\begin{equation}
m \frac{d^2 r}{dt^2} = -G \frac{M m}{r^2} - \frac{1}{2} C_d \rho(r) A(r) (v - v_{\text{sw}}(r)) |v - v_{\text{sw}}(r)|
\end{equation}

where $\frac{d^2r}{dt^2} = v\frac{dv}{dr}$. Therefore:

\begin{equation}
\boxed{v \frac{dv}{dr} = - \frac{G M_s}{r^2} - \frac{C_d \rho(r) A(r)}{2m} (v - v_{\text{sw}}(r)) |v - v_{\text{sw}}(r)|}
\label{dynamicsCME}
\end{equation}

\subsubsection{Environment Variables}

Here, to define the solar wind velocity, $v_{\text{sw}}(r)$, we employ Parker’s isothermal model \citep{Priest14}:  

\begin{equation}
    \frac{v_{\text{sw}}^2}{c_s^2} - \ln\left(\frac{v_{\text{sw}}^2}{c_s^2}\right) = 4 \ln\left(\frac{r}{r_c}\right) + 4 \frac{r_c}{r} - 3
    \label{eq:parker_isothermal}
\end{equation}

where:
\begin{itemize}
    \item \( c_s = \sqrt{\frac{k T}{m}} \): the isothermal sound speed, with \( k \) denoting Boltzmann’s constant, \( T \) the coronal temperature, and \( m \) the mean particle mass (typically the proton mass, \( m_p \)),
    \item \( r_c = \frac{G M_s}{2 c_s^2} \): the critical radius, i.e., the location where \( v_{\text{sw}} = c_s \) (transition from subsonic to supersonic flow), with \( G \) as the gravitational constant and \( M_s \) the solar mass.
\end{itemize} 

In addition, we must also specify the density profile in the interplanetary medium, $\rho(r)$, which is obtained from the mass continuity equation for a radial spherical outflow \citep{priest_magnetohydrodynamics_2014}:  

\begin{equation}
    \rho(r) = \frac{\dot{M}}{4\pi r^2 v_{\text{sw}}(r)}
\end{equation}

It is worth noting that this model is not particularly accurate at distances very close to the Sun, but it provides a reliable approximation throughout most of the interplanetary medium.

\subsubsection{CME Mass and Projected Area}
\label{Mass_and_ProjectedArea}

Once the parameters of the interplanetary medium have been established, it is necessary to specify the characteristic properties of the CME itself, such as its mass $m$ and its wetted area $A(r)$, i.e., the projected area on the plane perpendicular to its propagation direction.

To properly account for the global orientation of the structure, we apply a rotation with respect to the $x$-axis by an angle $\theta_x$, aligning the torus with the locally determined axis of propagation. Under this transformation, the surface of the Tapered Torus can be expressed as  

\[
\begin{cases}
x_{\text{rot}} = x = (R + a_{\text{max}} \cos \phi \cos \theta)\,\cos \phi, \\[6pt]
y_{\text{rot}} = y \cos \theta_x - z \sin \theta_x = (R + a_{\text{max}} \cos \phi \cos \theta)\,\sin \phi \cos \theta_x - \delta \, a_{\text{max}} \cos \phi \sin \theta \sin \theta_x, \\[6pt]
z_{\text{rot}} = y \sin \theta_x + z \cos \theta_x = (R + a_{\text{max}} \cos \phi \cos \theta)\,\sin \phi \sin \theta_x + \delta \, a_{\text{max}} \cos \phi \sin \theta \cos \theta_x.
\end{cases}
\]

In this parameterization:
\begin{itemize}
    \item $R$ is the major radius of the structure, i.e., the radius of the circular arc along which the elliptical cross-sections are swept to generate the croissant-shaped torus,
    \item $a(\phi)$ and $b(\phi)$ are the semi-major and semi-minor axes of the elliptical cross-section, related through $b(\phi) = \delta \, a(\phi)$, with $0 < \delta \leq 1$,
    \item $a_{\text{max}}$ denotes the maximum value of the semi-major axis, which occurs at $\phi = 0$, corresponding to the central part of the croissant-shaped torus,
    \item $\theta \in [0, 2\pi)$ is the poloidal angle, measuring the position around the elliptical cross-section of the torus,
    \item $\phi \in [-\pi/2, \pi/2]$ is the toroidal angle, restricted to half a revolution in order to describe the croissant-shaped geometry,
    \item $\theta_x$ is the rotation angle applied with respect to the $x$-axis to match the local CME orientation.
\end{itemize}

\begin{figure}[!ht]
    \centering
    \includegraphics[width=0.99\linewidth]{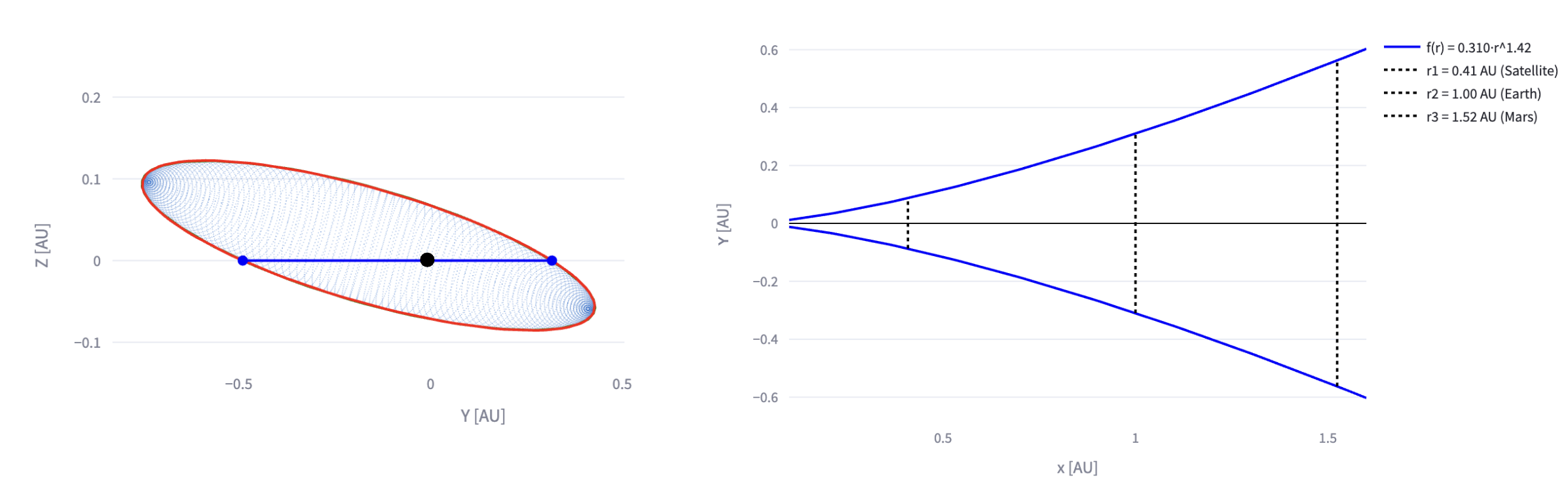}
    \caption{a) Wetted area of the Tapered Torus, showing the intersection with the $z=0$ plane and the spacecraft located along the $x$-axis. b) Swept area of the CME in the ecliptic plane.}
    \label{Wet_area_and_SweptArea}
\end{figure}

To compute the wetted area, the surface must be projected onto the plane perpendicular to the CME propagation direction. This projection can be carried out numerically, and the resulting contour can be well approximated by an ellipse (as illustrated in Figure \ref{Wet_area_and_SweptArea}a). Considering that the semi-major axis of the cross-section grows exponentially, the wetted area of the CME can then be expressed as:

\begin{equation}
    a(r) = a_0 \left( \frac{r}{r_0} \right)^{\alpha} \hspace{2cm} A(r) = \pi \delta a(r)^2 = \pi \delta a_0^2 \left( \frac{r}{r_0} \right)^{2\alpha}
\end{equation}

Since there is no way to determine the initial size of the CME, we will consider a standard initial value of \( a_0 \approx 3000 \, \text{km} \) at a distance of 2 solar radii, and we will seek the growth factor \( \alpha \), using the experimental values of \( b_1 \) known at \( r_1 \) and \( b_0 \) known at \( r_0 \):

\begin{equation}
    \alpha = \frac{\ln(b_1 / b_0)}{\ln(r_1 / r_0)}
\end{equation}

Finally, to compute the total volume we sweep an elliptical cross-section along 
\(\phi \in (-\pi/2, \pi/2)\). The area of a cross-section is given by
\[
A(\phi) = \pi\,a\,b\,\cos^2\phi,
\]
and integrating it along the central path, with differential length \(R\,d\phi\), yields
\begin{equation}
    V =  \frac{\pi^2 R a b}{2} = \frac{\pi^2 R \delta a^2}{2}.
\end{equation}

This expression corresponds to the CME volume at the time of observation. Then, by combining it with the average in-situ density of the CME, the total mass can be computed as
\[
m_{\mathrm{CME}} = \langle \rho \rangle \, V_{\mathrm{CME}},
\]
which, as previously discussed, is assumed to remain constant during the propagation.  

\subsubsection{Solving for the Velocity}

\begin{figure}[!ht]
    \centering
    \includegraphics[width=0.7\linewidth]{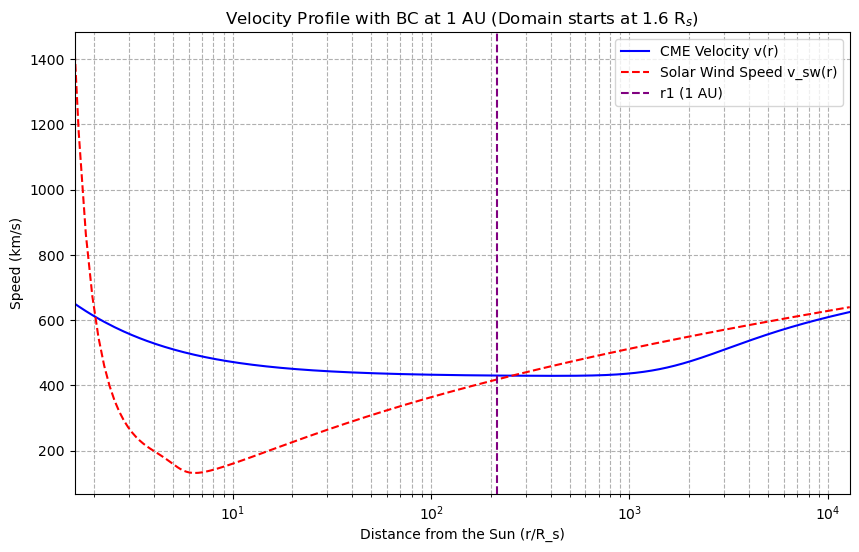}
    \caption{Propagation velocity of the CME and solar wind velocity.}
    \label{fig:PropagationVelocity}
\end{figure}

With all the above definitions, we can now solve Eq.~\ref{dynamicsCME}, obtaining the CME velocity as a function of its propagation through the interplanetary medium. By numerically integrating the ODE, we find the velocity profile of the CME along its propagation through the Solar System, as shown in Figure \ref{fig:PropagationVelocity}. This result agrees with experimental values of CME velocities at the moment they are ejected from the Sun, with values ranging between 500 km/s and 1000 km/s \citep{Nieves23, Schwenn06}.


Thus, once the velocity profile along the CME trajectory is determined, it can be integrated to obtain the travel time from the point of ejection at the Sun to the observational position. This procedure allows us to infer the launch time of the CME and to assess the accuracy of emission-time predictions.

The total travel time \( t \) from the initial position \( r_0 \) to the observed position \( r_1 \) is given by the integral of the inverse velocity profile \( v(r) \):

\begin{equation}
    t = \int_{r_0}^{r_1} \frac{dr}{v(r)}.
\end{equation}

This formulation also enables the prediction of the time at which the CME will cross a given heliocentric distance along its trajectory, as will be shown later for the cases of Earth’s and Mars’ orbits.

\section{Swept Area in the Ecliptic Plane}

An essential step in the study of CME propagation is the determination of the portion of the CME that intersects the ecliptic plane, as this is critical for assessing potential impacts on planetary bodies or artificial satellites.  

To this end, we first identify the intersection of the projected CME surface with the plane \(z=0\). Subsequently, we compute the area swept out by the CME in the ecliptic plane as it propagates.  

At the intersection with the ecliptic plane, the spacecraft is located at the position where $y=0$, i.e., along the $x$-axis.  

By measuring the lengths of the intersection on either side of the $x$-axis and knowing the heliocentric distance at which the CME is observed ($r_1$), we can estimate the correction angle required to determine the true propagation direction of the CME cut in the ecliptic plane. This correction angle is defined as  

\begin{equation}
    \varphi_{\mathrm{correction}} = \arcsin\!\left(\frac{\Delta L_{\mathrm{cut}}}{2\,r_1}\right),
\end{equation}

where \(\Delta L_{\mathrm{cut}}\) represents the difference in intersection lengths between the right and left sides. This correction provides the effective propagation direction of the central axis of the CME cut within the ecliptic plane.  

The intersection length is assumed to grow following a power-law profile attenuated by the parameter \(\alpha\), previously determined, such that  

\[
f(r) = f_0 \left(\frac{r}{r_0}\right)^{\alpha}.
\]

From this relation, the angular extent covered by the CME at a given distance can be expressed as  

\begin{equation}
    \gamma(r) = \arctan\left(\frac{f(r)}{r_{2}}\right),
\end{equation}

where $f(r)$ denotes the intersection length defined previously and $r_{2}$ is the heliocentric distance of interest.  

Finally, we can assess whether the swept area intersects planetary bodies or artificial satellites by predicting the time at which the CME reaches their orbital positions and determining if these bodies are contained within the angular extent $\gamma(r)$. For this purpose, the implementation in the \href{https://xcme-nasa.streamlit.app/}{X-CME} code makes use of the \texttt{Astropy} and \texttt{SunPy} libraries, together with the \texttt{heliosat} package, which provide access to NASA JPL Horizons ephemerides and heliospheric coordinate transformations. These tools enable the retrieval of planetary and spacecraft positions at the predicted CME crossing times.  

\begin{figure}[!ht]
    \centering
    \includegraphics[width=0.6\linewidth]{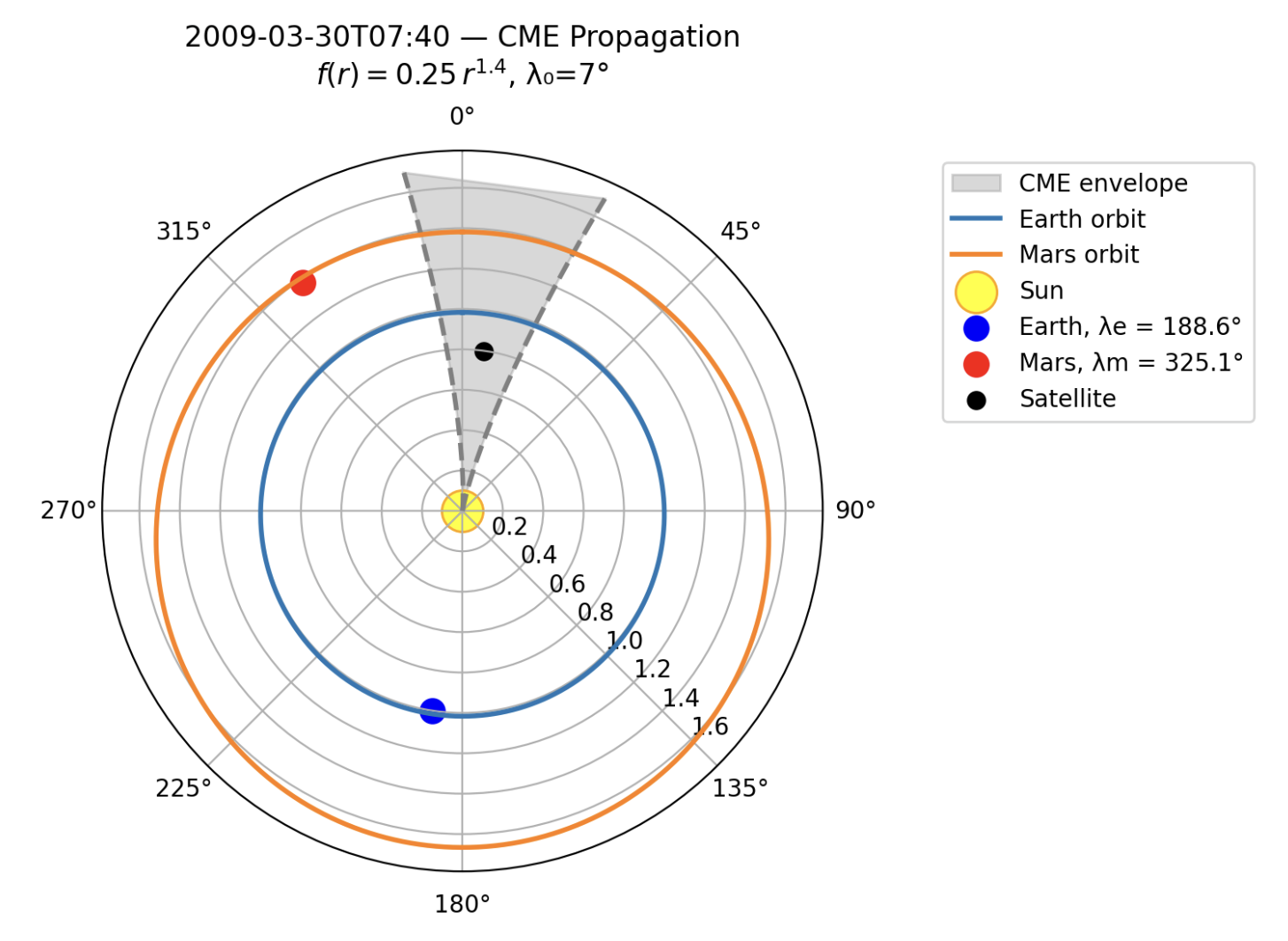}
    \caption{CME angular extent projected onto the Solar System ecliptic plane.}
    \label{fig:enter-label}
\end{figure}

An example of the CME sweep in the inner heliosphere is shown in Fig.~\ref{fig:enter-label}.  

\section{Analysis of Real Events}
\label{S-RealEvents}

In this section we validate the CME–propagation framework described in the previous chapters by applying it to two real heliospheric events. In both cases, the internal magnetic structure of the flux rope is reconstructed at an intermediate heliocentric distance and then propagated through interplanetary space using the dynamical model of Section~\ref{dynamicsCME}. The predicted arrival times and angular extents at Earth are finally compared with in–situ measurements from the \emph{WIND} spacecraft at L1.

For the present validation we focus exclusively on the \emph{radial–poloidal} reconstruction introduced in our previous work \citep{MasoMoreno2025prep}. This model provides a more realistic description of the helical and cross–sectional properties of the magnetic flux rope, while the global CME evolution---mass, expansion and interaction with the background solar wind---is treated in exactly the same way as in the synthetic tests of Sections~\ref{Mass_and_ProjectedArea} and \ref{dynamicsCME}. The only event–to–event differences therefore arise from the local magnetic–field fits and the resulting geometrical parameters of the embedded Tapered Torus.

\begin{figure}[!ht]
    \centering
    \begin{minipage}{0.48\textwidth}
        \centering
        \includegraphics[width=\linewidth]{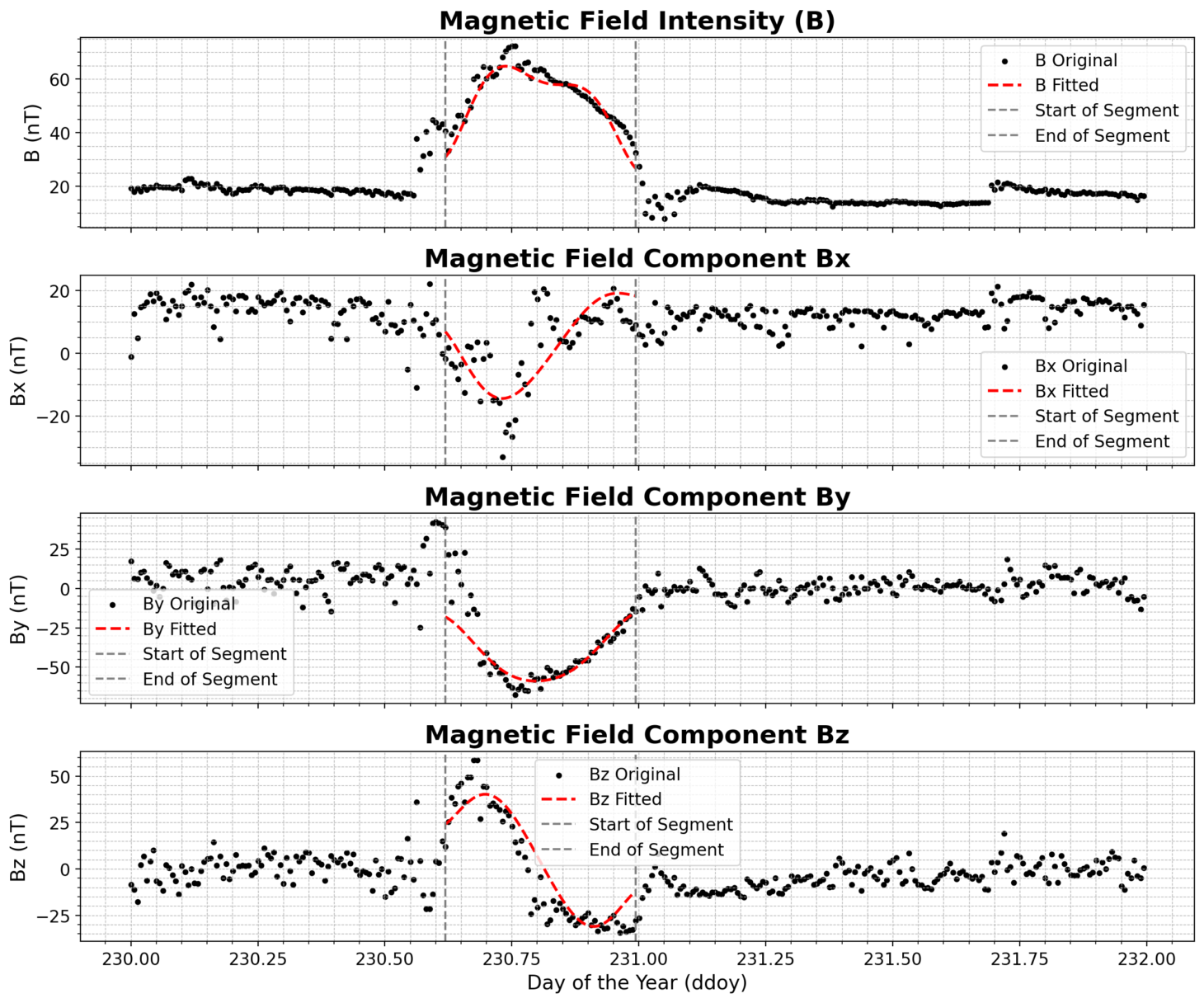}
        \caption*{Event 1: PSP (18 August 2022).}
    \end{minipage}
    \hfill
    \begin{minipage}{0.48\textwidth}
        \centering
        \includegraphics[width=\linewidth]{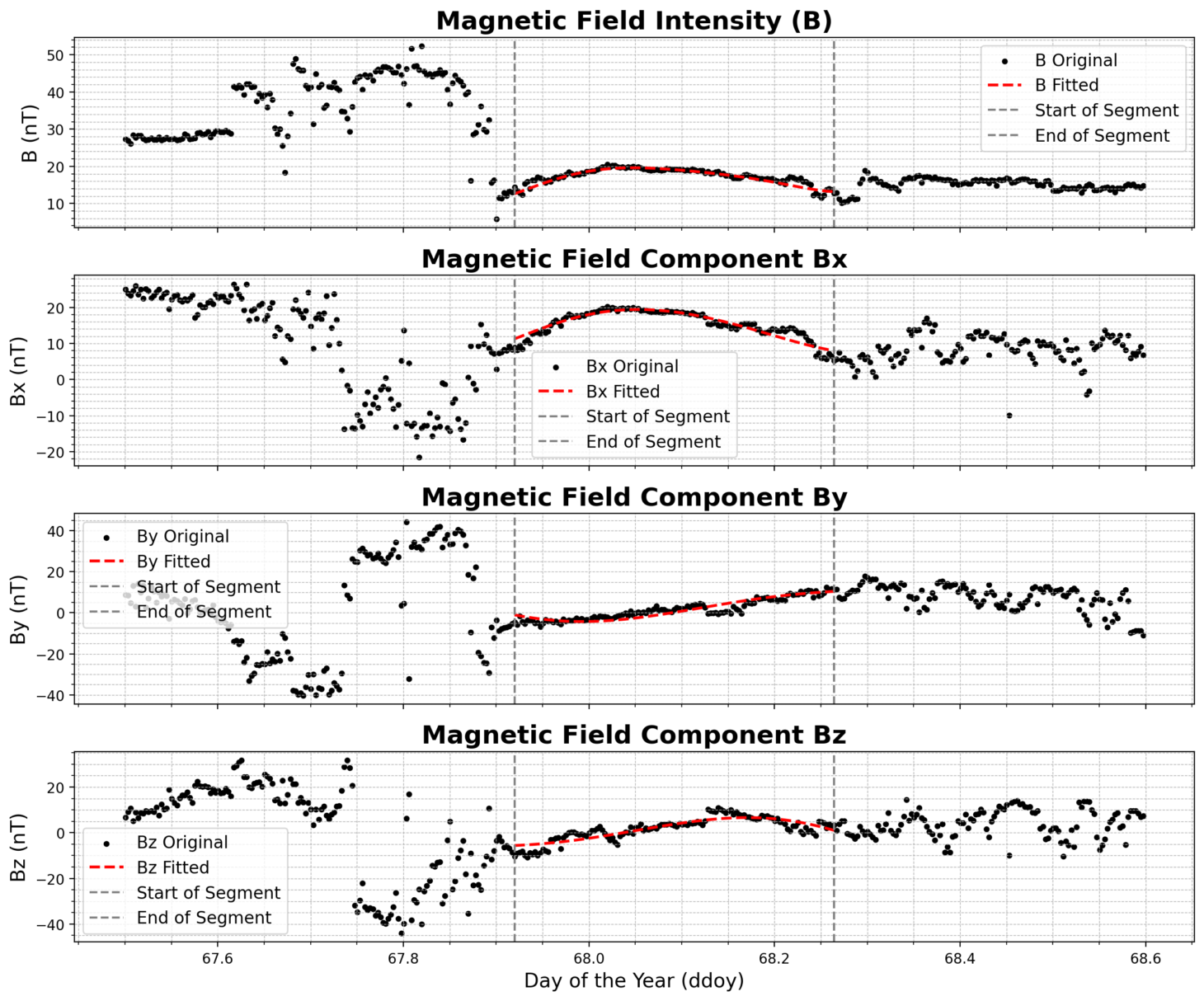}
        \caption*{Event 2: SolO (8 March 2022).}
    \end{minipage}
    \caption{Magnetic–field fits obtained with the radial–poloidal model for the two validation events. In both cases the model reproduces the characteristic rotation of the magnetic field and the enhanced field magnitude within the flux–rope interval.}
    \label{fig:real-events-fits}
\end{figure}

The two selected CMEs correspond to a flux–rope interval observed by \emph{Parker Solar Probe} (PSP) on 18 August 2022 (Event~1) and another interval detected by \emph{Solar Orbiter} (SolO) on 8 March 2022 (Event~2). In both events the in–situ signatures include a smooth rotation of the magnetic field, enhanced $|B|$, and reduced proton temperature, consistent with a flux–rope passage. The fitted parameters of the radial–poloidal model are then used as input for the global X-CME propagation scheme.

\newcolumntype{Y}{>{\centering\arraybackslash}X}

\begin{table}[H]
    \centering
    \begin{tabularx}{\textwidth}{l Y Y}
        \toprule
        \textbf{Parameter} & \textbf{Event 1 (PSP, 18 Aug 2022)} & \textbf{Event 2 (SolO, 8 Mar 2022)} \\
        \midrule
        \multicolumn{3}{l}{\textbf{Key timestamps}} \\
        \midrule
        Launch time from the Sun [UT]             & 2022-08-17 T09{:}04 & 2022-03-07 T03{:}59 \\
        CME detection at spacecraft [UT]          & 2022-08-18 T15{:}21 & 2022-03-08 T22{:}04 \\
        Predicted arrival at Earth's orbit [UT]   & 2022-08-19 T17{:}50 & 2022-03-10 T23{:}26 \\
        Predicted arrival at Mars' orbit [UT]     & 2022-08-21 T02{:}20 & 2022-03-13 T00{:}16 \\
        \midrule
        \multicolumn{3}{l}{\textbf{Satellite and planetary positions at their respective timestamps}} \\
        \midrule
        Spacecraft distance $r_1$ [AU], longitude $\gamma_1$ [deg] 
            & $(0.554,\; 21.25^{\circ})$ & $(0.477,\; 3.00^{\circ})$ \\
        Earth distance $r_2$ [AU], longitude $\gamma_2$ [deg] 
            & $(1.012,\; 0.00^{\circ})$ & $(0.993,\; 0.00^{\circ})$ \\
        Mars distance $r_3$ [AU], longitude $\gamma_3$ [deg] 
            & $(1.407,\; 46.44^{\circ})$ & $(1.447,\; 102.29^{\circ})$ \\
        \midrule
        \multicolumn{3}{l}{\textbf{CME propagation details (radial–poloidal model)}} \\
        \midrule
        Ecliptic longitude of CME center [deg]   & $18.10^{\circ}$ & $3.16^{\circ}$ \\
        Flux–rope inclination [deg]             & $-27.48^{\circ}$ & $-35.20^{\circ}$ \\
        Angular width at Earth's distance [deg] & $(4.13^{\circ},\, 32.07^{\circ})$  & $(349.74^{\circ},\, 16.59^{\circ})$ \\
        Angular width at Mars' distance [deg]   & $(0.69^{\circ},\, 35.52^{\circ})$  & $(348.61^{\circ},\, 17.72^{\circ})$ \\
        Encounter with Earth                    & No  & Yes \\
        Encounter with Mars                     & No  & No  \\
        \bottomrule
    \end{tabularx}
    \caption{Key timestamps, multi–point geometry, and propagation properties for the two validation events, as derived from the radial–poloidal reconstruction and the X-CME propagation scheme.}
    \label{tab:real-events-comparison}
\end{table}

The radial–poloidal fits produce robust large–scale geometries in both events: axis orientations and cross–sectional sizes are well constrained by the in–situ data, while the background solar wind and CME expansion are treated identically in the two cases. The main differences arise in the flux–rope inclination and ecliptic longitude, which control whether the propagated torus intersects the orbits of Earth and Mars.

\begin{figure}[!ht]
    \centering
    \begin{minipage}{0.43\textwidth}
        \centering
        \includegraphics[width=\linewidth]{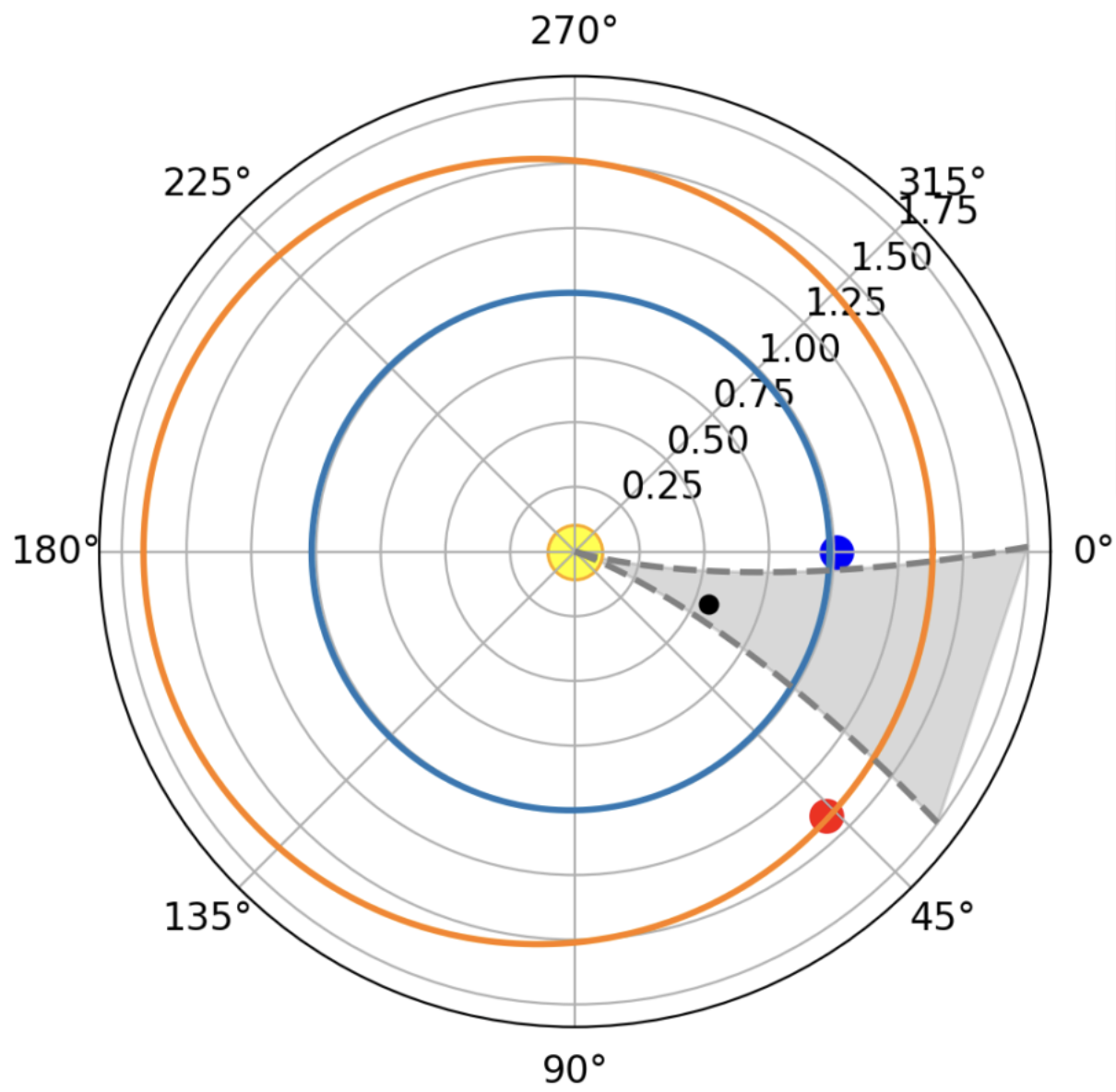}
        \caption*{Event 1: propagation forecast from PSP to Earth and Mars.}
    \end{minipage}
    \hfill
    \begin{minipage}{0.43\textwidth}
        \centering
        \includegraphics[width=\linewidth]{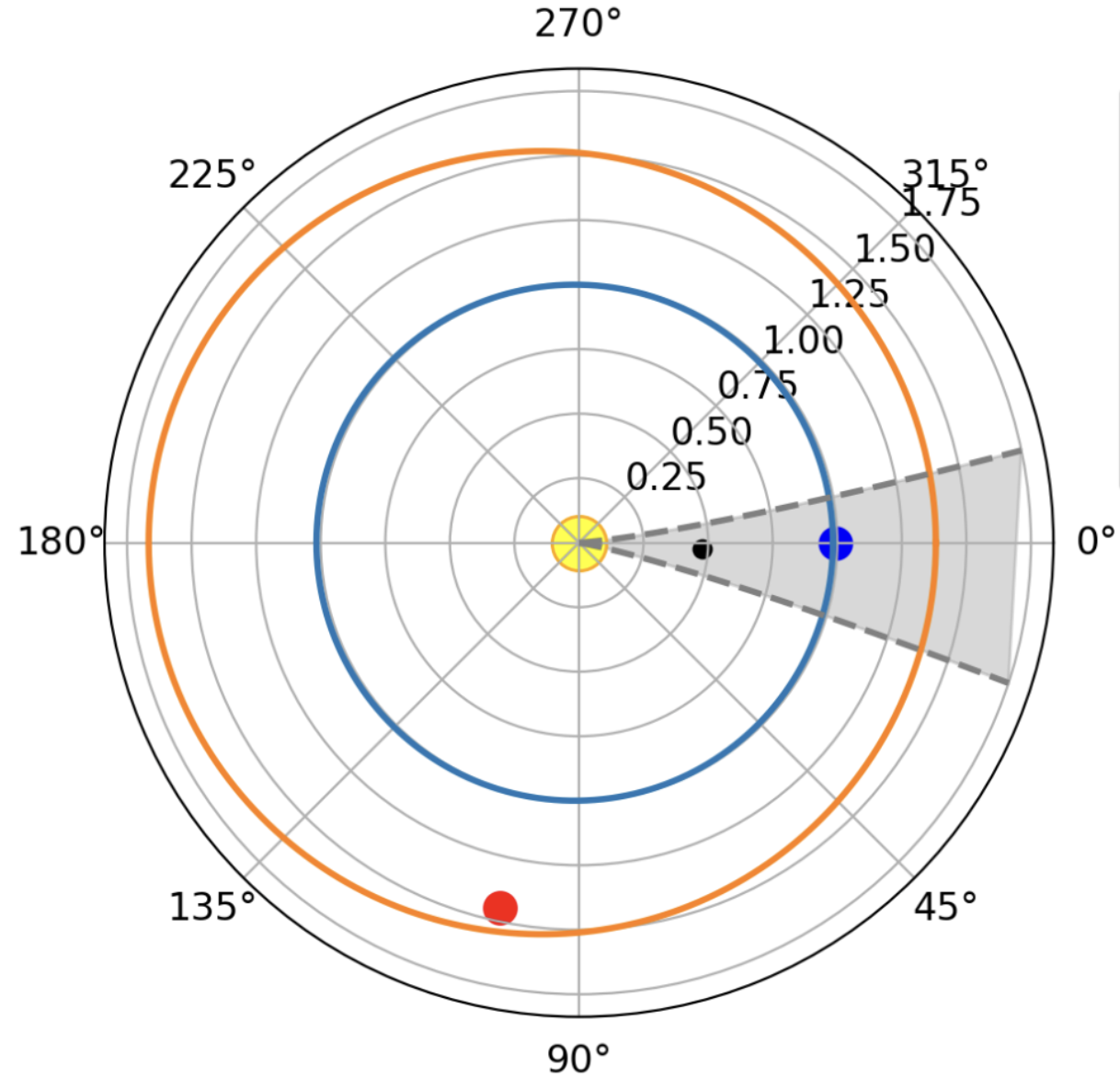}
        \caption*{Event 2: propagation forecast from SolO to Earth and Mars.}
    \end{minipage}
    \caption{CME propagation in the ecliptic plane for the two validation events, using the radial–poloidal reconstruction as input to the X-CME model. The shaded regions represent the intersection of the Tapered Torus with the ecliptic plane at successive times.}
    \label{fig:real-events-propagation}
\end{figure}

For PSP Event~1, the inferred inclination is such that the CME propagates between the orbits of Earth and Mars, with only the outermost flank approaching Earth's longitude. We note that, throughout this section, longitudes are expressed in the Heliocentric Earth Ecliptic (HEE) coordinate system, in which the Earth is located by convention at $0^{\circ}$ longitude at all times. Under this reference system, the model predicts a near miss or, at most, a glancing encounter. For SolO Event~2, the CME center remains much closer to the ecliptic and to Earth's longitude, leading to a frontal impact prediction with no significant intersection with Mars' orbit.

\begin{figure}[!ht]
    \centering
    \begin{minipage}{0.48\textwidth}
        \centering
        \includegraphics[width=\linewidth]{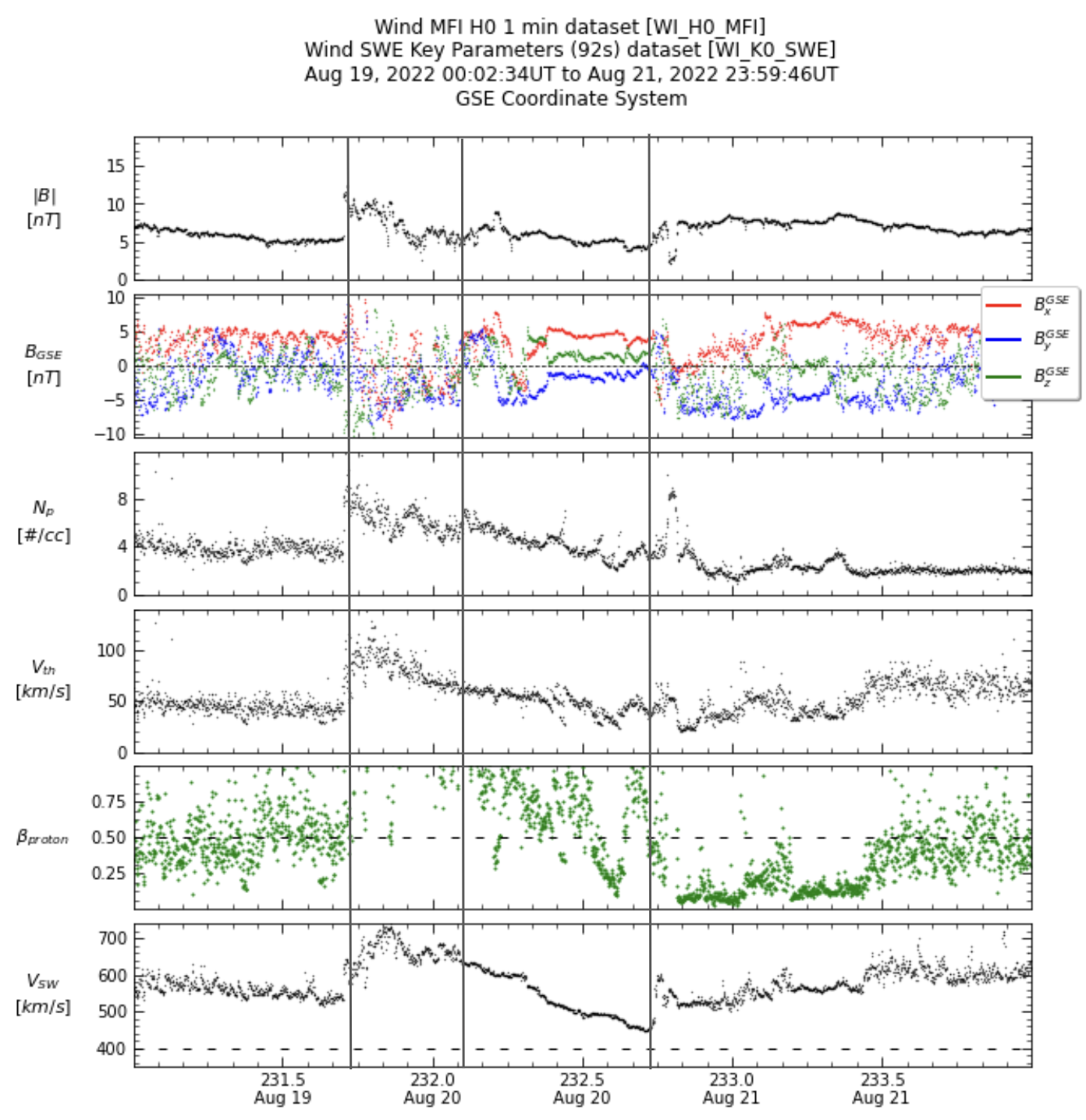}
        \caption*{Event 1: \emph{WIND} magnetic and plasma data on 19-21 August 2022.}
    \end{minipage}
    \hfill
    \begin{minipage}{0.48\textwidth}
        \centering
        \includegraphics[width=\linewidth]{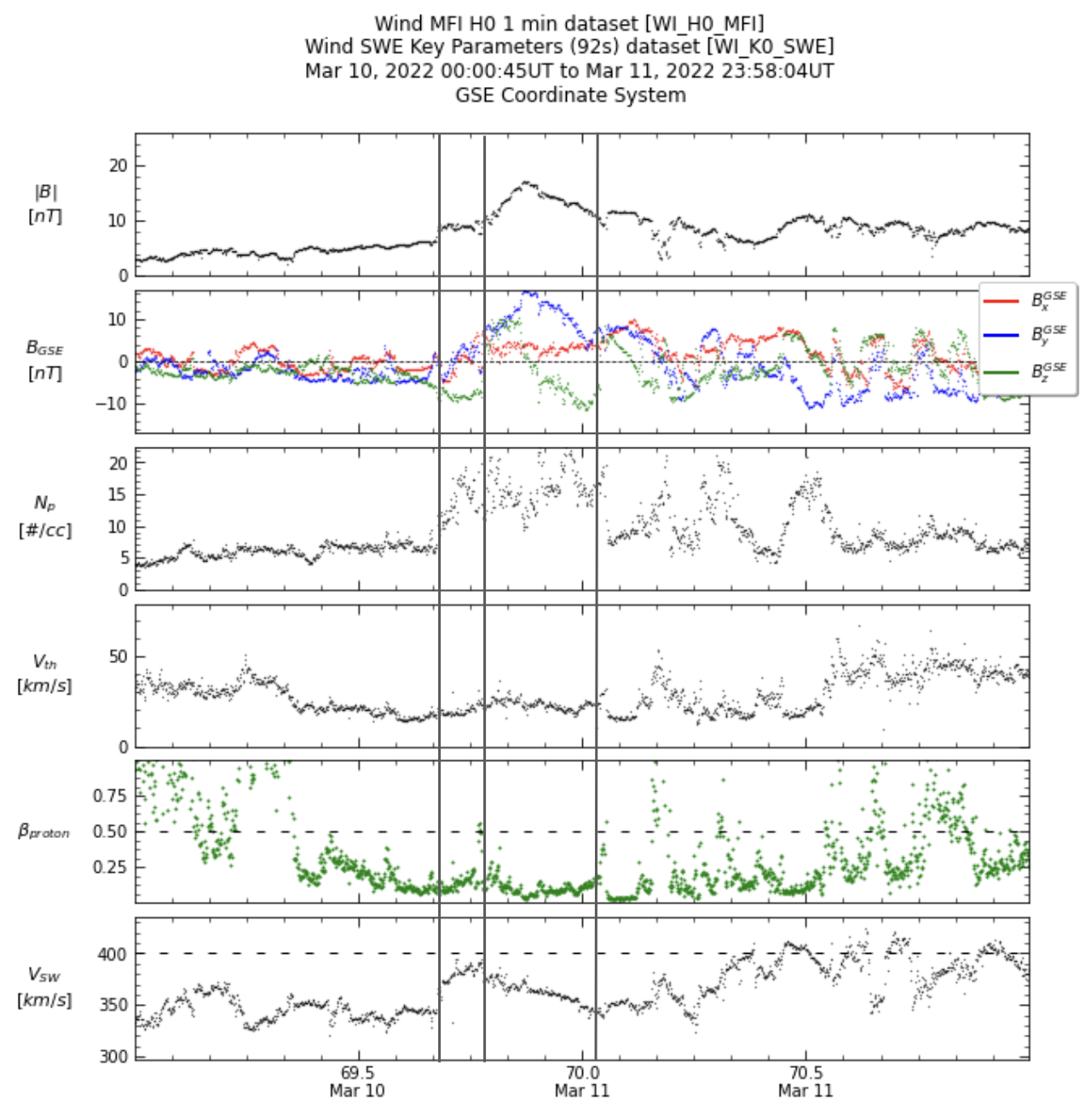}
        \caption*{Event 2: \emph{WIND} data for 9--12 March 2022.}
    \end{minipage}
    \caption{In–situ measurements at L1 from the \emph{WIND} spacecraft around the predicted arrival times of the two CMEs, retrieved from the LASSOS catalog~\citep{lassos_catalogs}. Both intervals display characteristic ICME signatures, with different impact geometries.}
    \label{fig:real-events-wind}
\end{figure}

In Event~1, the \emph{WIND} data show a clear shock followed by a modest enhancement in $|B|$ and a partially organized rotation of the field components, consistent with a \emph{glancing blow} in which the spacecraft traverses the CME flank or its surrounding sheath. The predicted arrival time from the X-CME model differs from the observed shock time by less than $\sim 2$~hours, highlighting the high temporal accuracy achievable when the model is anchored to an intermediate–distance spacecraft.

In Event~2, the \emph{WIND} measurements reveal a well–defined flux–rope interval with enhanced $|B|$, low proton $\beta_p$, and a smooth rotation of the magnetic field, indicative of a near–central crossing. The model predicts the arrival of the CME at Earth's orbit to within $\sim 4$~hours of the observed onset time, significantly better than typical coronagraph–based forecasts. The remaining discrepancy can be attributed to uncertainties in the initial positioning of the fitted flux–rope segment relative to the global CME front, as well as to the simplified treatment of the background solar wind and CME expansion.

Overall, the two validation events demonstrate that combining radial–poloidal flux–rope reconstructions with the X-CME propagation framework can deliver CME arrival–time predictions at Earth with errors on the order of a few hours (typically $\sim$4--6~h), while also capturing the correct impact geometry (central encounter versus glancing blow or near miss). At the same time, these results emphasize the sensitivity of impact/no–impact forecasts to the inferred flux–rope inclination and underline the need for future extensions of the model to include the sheath region explicitly.

\bibliographystyle{spr-mp-sola}
\bibliography{references}  

@ARTICLE{2018ARA&A..56..135Z,
  author       = {Zurbuchen, T. H. and Richardson, I. G.},
  title        = {In-Situ Solar Wind and Magnetic Field Measurements: A Necessary Requirement for Understanding CMEs and Space Weather},
  journal      = {\araa},
  year         = {2018},
  month        = sep,
  volume       = {56},
  pages        = {135--160},
  doi          = {10.1146/annurev-astro-081817-051839}
}

@MISC{lassos_catalogs,
  author       = {{NASA Goddard Space Flight Center}},
  title        = "{LASSOS Catalogs}",
  howpublished = {\url{https://science.gsfc.nasa.gov/lassos/catalogs.shtml}},
  note         = {Accessed: 2025-05-12},
  year         = {2025}
}

@MISC{MasoMoreno2025prep,
  author = {Massó-Moreno, M. et al.},
  title  = "{Generalization of Elliptical--Cylindrical Flux Rope Models for ICME Reconstruction}",
  note   = {In preparation},
  year   = {2025}
}

@ARTICLE{Nieves23,
   author = {{Nieves-Chinchilla}, T. and {Pal}, S. and {Salman}, T.~M. and {Carcaboso}, F. and {Guidoni}, S.~E. and {Cremades}, H. and {Narock}, A. and {Balmaceda}, L.~A. and {Lynch}, B.~J. and {Al-Haddad}, N. and {Rodr{\'\i}guez-Garc{\'\i}a}, L. and {Narock}, T.~W. and {Dos Santos}, L.~F.~G. and {Regnault}, F. and {Kay}, C. and {Winslow}, R.~M. and {Palmerio}, E. and {Davies}, E.~E. and {Scolini}, C. and {Weiss}, A.~J. and {Alzate}, N. and {Jeunon}, M. and {Pujadas}, R.},
    title = "{Redefining Flux Ropes in Heliophysics}",
  journal = {Front. Astron. Space Sci.},
     year = 2023,
    month = mar,
   volume = 10,
    pages = {56},
      doi = {10.3389/fspas.2023.1114838},
   adsurl = {2023FrASS..1014838N}
}

@ARTICLE{Schwenn06,
   author = {{Schwenn}, R.},
    title = "{Space Weather: The Solar Perspective}",
  journal = {Living Rev. Sol. Phys.},
     year = 2006,
    month = dec,
   volume = 3,
   number = 1,
    pages = {2},
      doi = {10.12942/lrsp-2006-2},
   adsurl = {2006LRSP....3....2S}
}

@BOOK{Priest14,
   author = {{Priest}, E.~R.},
    title = "{Magnetohydrodynamics of the Sun}",
publisher = {Cambridge University Press},
     year = 2014,
    month = mar,
      isbn = {978-0-521-85471-9},
   adsurl = {https://www.cambridge.org/9780521854719}
}

@ARTICLE{Nieves18,
   author = {{Nieves-Chinchilla}, T. and {Linton}, M.~G. and {Hidalgo}, M.~A. and {Vourlidas}, A.},
    title = "{Elliptic-Cylindrical Analytical Flux Rope Model for Magnetic Clouds}",
  journal = {\apj},
     year = 2018,
    month = jul,
   volume = 861,
   number = 1,
    pages = {139-153},
      doi = {10.3847/1538-4357/aac951},
   adsurl = {2018ApJ...861..139N}
}

@ARTICLE{Vrsnak10,
   author = {{Vršnak}, B. and {Žic}, T. and {Falkenberg}, T.~V. and {Möstl}, C. and {Vennerstrom}, S. and {Vrbanec}, D.},
    title = "{The role of aerodynamic drag in propagation of interplanetary coronal mass ejections}",
  journal = {A\&A},
     year = 2010,
    month = mar,
   volume = 512,
    pages = {A43},
      doi = {10.1051/0004-6361/200913482},
   adsurl = {2010A&A...512A..43V}
}

@ARTICLE{Mishra2021,
  author       = {{Mishra}, {Wageesh} and {Doshi}, {Urmi} and {Srivastava}, {Nandita}},
  title        = {Radial Sizes and Expansion Behavior of ICMEs in Solar Cycles 23 and 24},
  journal      = {Frontiers in Astronomy and Space Sciences},
  year         = {2021},
  volume       = {8},
  pages        = {713999},
  doi          = {10.3389/fspas.2021.713999}
}

@ARTICLE{Kay2015,
  author       = {Kay, C. and Opher, M. and Evans, R. M.},
  title        = {Global Trends of CME Deflections Based on CME and Solar Parameters},
  journal      = {\apj},
  year         = {2015},
  volume       = {805},
  number       = {2},
  pages        = {168},
  doi          = {10.1088/0004-637X/805/2/168}
}

@ARTICLE{Vourlidas2010,
  author       = {Vourlidas, A. and Howard, R. A. and Esfandiari, E. and Patsourakos, S. and Yashiro, S. and Michalek, G.},
  title        = {Comprehensive Analysis of Coronal Mass Ejection Mass and Energy over a Full Solar Cycle},
  journal      = {\apj},
  year         = {2010},
  volume       = {722},
  number       = {2},
  pages        = {1522-1538},
  doi          = {10.1088/0004-637X/722/2/1522}
}

@ARTICLE{Bein2013,
  author       = {Bein, B. M. and Temmer, M. and Vourlidas, A. and Veronig, A. M. and Utz, D.},
  title        = {The height evolution of the ``true'' CME mass derived from STEREO COR1 and COR2 observations},
  journal      = {\apj},
  year         = {2013},
  volume       = {768},
  number       = {1},
  pages        = {31},
  doi          = {10.1088/0004-637X/768/1/31}
}

@BOOK{MeyerVernet2007,
  author       = {Meyer-Vernet, N.},
  title        = {Basics of the Solar Wind},
  publisher    = {Cambridge University Press},
  year         = {2007},
  doi          = {10.1017/CBO9780511535765}
}

@BOOK{Chen2016,
  author       = {Chen, F. F.},
  title        = {Introduction to Plasma Physics and Controlled Fusion},
  publisher    = {Springer},
  year         = {2016},
  edition      = {3rd},
  doi          = {10.1007/978-3-319-22309-4}
}

@ARTICLE{2024SpWea..2203951K,
  author       = {Kay, C. and Palmerio, E. and Riley, P. and Chulaki, A. and {et al.}},
  title        = {Updating Measures of CME Arrival Time Errors},
  journal      = {Space Weather},
  year         = {2024},
  month        = jul,
  volume       = {22},
  eid          = {e2024SW003951},
  doi          = {10.1029/2024SW003951}
}

\end{document}